\documentclass{aa}
\usepackage{graphicx,epstopdf,titlesec,siunitx,amsmath,fixltx2e,subcaption,textgreek,amsmath,lineno,tabularx}
\usepackage[varg]{txfonts}
\bibpunct{(}{)}{;}{a}{}{,}

\begin{document}

\title{Detection of H\textsubscript{3}\textsuperscript{+} auroral emission in Jupiter's 5-micron window}

\author{R. S. Giles\inst{\ref{oxford}} \and L. N. Fletcher\inst{\ref{leicester}} \and P. G. J. Irwin\inst{\ref{oxford}} \and H. Melin\inst{\ref{leicester}} \and T. S. Stallard\inst{\ref{leicester}}}

\institute{Atmospheric, Oceanic \& Planetary Physics, Department of Physics, University of Oxford, Clarendon Laboratory, Parks Road, Oxford OX1 3PU, UK\label{oxford} \and Department of Physics and Astronomy, University of Leicester, University Road, Leicester LE1 7RH, UK\label{leicester}}

\abstract{

We use high-resolution ground-based observations from the VLT CRIRES instrument in December 2012 to identify sixteen previously undetected H\textsubscript{3}\textsuperscript{+} emission lines from Jupiter's ionosphere. These emission lines are located in Jupiter's 5-micron window (4.5-5.2 \textmu m), an optically-thin region of the planet's spectrum where the radiation mostly originates from the deep troposphere. The H\textsubscript{3}\textsuperscript{+} emission lines are so strong that they are visible even against this bright background. We measure the Doppler broadening of the H\textsubscript{3}\textsuperscript{+} emission lines in order to evaluate the kinetic temperature of the molecules, and we obtain a value of 1390$\pm$160 K. We also measure the relative intensities of lines in the \textnu\textsubscript{2} fundamental in order to calculate the rotational temperature, obtaining a value of 960$\pm$40 K. Finally, we use the detection of an emission line from the 2\textnu\textsubscript{2}(2)-\textnu\textsubscript{2} overtone to measure a vibrational temperature of 925$\pm$25 K. We use these three independent temperature estimates to discuss the thermodynamic equilibrium of Jupiter's ionosphere.

}

\maketitle

\section{Introduction}

Observations of H\textsubscript{3}\textsuperscript{+} lines are a valuable tool in studying Jupiter's upper atmosphere. It can be used to measure ionospheric temperatures, and to trace energy inputs from high-energy particles and solar radiation. The species itself also directly affects the ionospheric conditions, both acting as a stabilising `thermostat', and providing the main contribution to ionospheric conductivity. Further details on the role of H\textsubscript{3}\textsuperscript{+} in planetary atmospheres can be found in~\citet{miller06}.

The first spectroscopic detection of H\textsubscript{3}\textsuperscript{+} emission from Jupiter's ionosphere was made by~\citet{drossart89} in the 2\textnu\textsubscript{2}($l=2$) overtone band at 2 \textmu m in the K-band. This was almost immediately followed by detections of the \textnu\textsubscript{2} fundamental at 4 \textmu m in the L-band \citep[e.g.][]{oka90}, and more recently by detection of the 2\textnu\textsubscript{2}(0)-\textnu\textsubscript{2} (4 \textmu m) and 3\textnu\textsubscript{2}(3)-\textnu\textsubscript{2} (2 \textmu m) overtones~\citep{stallard02,raynaud04}. The strongest H\textsubscript{3}\textsuperscript{+} signature is localised around the northern and southern auroral ovals~\citep{drossart92}; in these regions of the planet, high-energy electrons are accelerated along the magnetic field lines into the upper atmosphere, where they ionise Jupiter's neutral gases to produce H\textsubscript{2}\textsuperscript{+}. H\textsubscript{2}\textsuperscript{+} and H\textsubscript{2} can then combine to produce H\textsubscript{3}\textsuperscript{+} and H, a reaction that is so efficient that very little H\textsubscript{2}\textsuperscript{+} remains in the upper atmosphere. In addition to this auroral effect, there is a planet-wide signal due to extreme ultra-violet radiation from the sun which also causes ionisation~\citep{miller97}.

Studying the H\textsubscript{3}\textsuperscript{+} emission lines can provide us with several different types of temperature measurements for Jupiter's ionosphere. The kinetic temperature, T\textsubscript{kin}, can be derived from the Doppler broadening of individual lines. The rotational temperature, T\textsubscript{rot}, can be derived by comparing the relative intensities of different rotational lines within the same vibrational manifold. The vibrational temperature, T\textsubscript{vib}, can be derived by simultaneously measuring lines from multiple vibrational levels. If the gas is in local thermodynamic equilibrium (LTE), then these three temperatures should be the same. The kinetic temperature of H\textsubscript{3}\textsuperscript{+} in Jupiter's ionosphere has been measure once before~\citep{drossart93}, the vibrational temperature has been measured twice~\citep{stallard02,raynaud04}, and several studies have measured the rotational temperature~\citep{drossart89, oka90, maillard90}.

In this work, we use high-resolution ground-based observations to identify previously undetected H\textsubscript{3}\textsuperscript{+} emission lines in the 4.5-5.2 \textmu m range, belonging to the \textnu\textsubscript{2} fundamental and to the 2\textnu\textsubscript{2}(2)-\textnu\textsubscript{2} overtone. These emission lines are located in Jupiter's 5-micron atmospheric window; at these wavelengths, Jupiter's atmosphere is optically thin, allowing us to view bright radiation from deep in the planet's troposphere. We then measure the broadening of these lines to evaluate T\textsubscript{kin} and we measure the relative intensities of the \textnu\textsubscript{2} lines to evaluate T\textsubscript{rot}. Finally, we compare the fundamental and overtone lines to evaluate T\textsubscript{vib}. This is first time that all three temperatures have been measured simultaneously. 

\section{Observations and data reduction}

Observations of Jupiter were made using the CRIRES instrument at the European Southern Observatory's Very Large Telescope (VLT). These observations took place on 12 November 2012 at 05:00-05:40 UT. CRIRES is a cryogenic high-resolution infrared echelle spectrograph which provides long-slit spectroscopy with a resolving power of 96,000~\citep{kaufl04}. The observations were made in 14 different wavelength settings that together cover the entirety of Jupiter's 5-micron atmospheric window. Each wavelength setting was observed at a slightly different time, so corresponds to a slightly different longitude, ranging from 87$^{\circ}$W to 108$^{\circ}$W. The 0.2$\times$40'' slit was aligned north-south along Jupiter’s central meridian; on this date, Jupiter's angular diameter was 48'' so the observations cover the north pole, but not the south pole. Observations were also made of a standard star, HIP 22509 (A1 Vn type, RA: 04:50:36.72, Dec: 08:54:00.56).

The initial data reduction was performed using the EsoRex pipeline~\citep{ballester06}. This included bias-subtraction, flat-fielding, wavelength calibration using known telluric lines, and noise calculation using sky observations. Corrections were subsequently performed to account for tilting in both the spectral and spatial directions. In the spectral direction, this was achieved using telluric lines, and in the spatial direction, this was achieved by cross-correlating the latitudinal profile at different wavelengths. The jovian spectra were divided through by the standard star observations in order to account for telluric absorption and to provide radiometric calibration. The estimated flux loss due to the narrow slit (0.2'' compared to a seeing of 0.4'') was accounted for in this calibration. We find that the 5-micron brightness temperatures vary between 200 K in the coolest parts of the planet and 260 K in the brightest regions. This is roughly consistent with the Cassini VIMS 5-\textmu m observations, which showed a 190-240 K range~\citep{giles15}; any differences are likely to be due to how centrally the standard star was located within the narrow slit. It should be noted that the absolute calibration does not impact the quantitative results in this paper, as we only measure the widths and relative intensities of the emission lines.

Geometric data were calculated using the angular size of the planet, the angular size of each pixel and the location of the planet's limb. We assumed that the slit was aligned with Jupiter's central meridian.

\section{Analysis}

\subsection{Line identification}

We detected sixteen H\textsubscript{3}\textsuperscript{+} emission lines in this spectral region. These lines were identified using the spectroscopic line list of ~\citet{kao91}, and are listed in Table~\ref{tab:emission_features}. Fifteen of the lines belong to the \textnu\textsubscript{2} fundamental band, and the remaining line (located at 4.6118 \textmu m) belongs to the 2\textnu\textsubscript{2}(2)-\textnu\textsubscript{2} overtone. This is the first time that an emission line from this overtone has been observed in Jupiter's atmosphere. All of the lines are in the P-branch of the vibrational manifold, and the line identification describes the (J,K,l) quantum numbers of the lower state.

\begin{table}
\begin{tabularx}{\textwidth/2}{>{\raggedright\arraybackslash}X  >{\raggedright\arraybackslash}X  >{\raggedright\arraybackslash}X >{\raggedright\arraybackslash}X}
\hline
Line centre (\textmu m) & Line centre (cm\textsuperscript{-1}) & Line identification & Doppler broadening \space (10\textsuperscript{-5} \textmu m) \\
\hline
4.59605 & 2175.78 & P(5,0) & 8.2$\pm$0.7\\
4.60232 & 2172.82 & P(5,1) & 7.5$\pm$1.9\\
4.61180 & 2168.35 & P(5,6+) (overtone) & 8.7$\pm$1.4\\
4.62048 & 2164.28 & P(5,2) & 8.2$\pm$0.6\\
4.64493 & 2152.89 & P(5,3) & 6.5$\pm$0.2\\
4.67213 & 2140.35 & P(5,4) & 7.1$\pm$0.2\\
4.48401 & 2134.92 & P(5,5) & 7.6$\pm$0.2\\
4.71140 & 2122.51 & P(6,1) & 6.9$\pm$1.4\\
4.73206 & 2113.24 & P(6,2) & 6.7$\pm$1.3\\
4.76956 & 2096.63 & P(6,3) & 7.4$\pm$0.6\\
4.80901 & 2079.43 & P(6,4) & 6.9$\pm$0.5\\
4.87446 & 2051.51 & P(6,6) & 9.3$\pm$0.2\\
4.91807 & 2033.32 & P(7,4) & 9.2$\pm$1.6\\
4.98184 & 2007.29 & P(7,5) & 8.2$\pm$1.2\\
5.04318 & 1982.87 & P(7,6) & 10.2$\pm$0.5\\
5.16608 & 1935.71 & P(8,6) & 9.7$\pm$1.8\\
\end{tabularx}
\caption{Identified H\textsubscript{3}\textsuperscript{+} emission features in Jupiter's 5-micron window.}
\label{tab:emission_features}
\end{table}

The first emission line to be identified was the P(5,5) line at 4.48401 \textmu m. Using this emission line, we explored how the strength of the H\textsubscript{3}\textsuperscript{+} emission varied with latitude. In order to improve the signal-to-noise, the data was smoothed with a spatial bin width of three pixels. At each pixel, we used the IDL MPFIT routine~\citep{markwardt09} to fit the P(5,5) emission line. As CRIRES is an echelle spectrometer, the instrumental line shape is triangular, and the resolving power of 96,000 translates into a resolution of ${\sim}5{\times}10^{-5}$ \textmu m. Because the slit is very narrow, the broadening that occurs from Jupiter's rotation and the horizontal flow of gas is expected to be negligible. The function used to fit the data was therefore a convolution of a known triangular lineshape (from the instrument) and an unknown Gaussian curve (from the Doppler broadening). There were six variables used: three for the Gaussian component (line centre, line width, line strength) and three for the background (a quadratic fit).

\begin{figure}
\centering
\includegraphics[width=9.2cm]{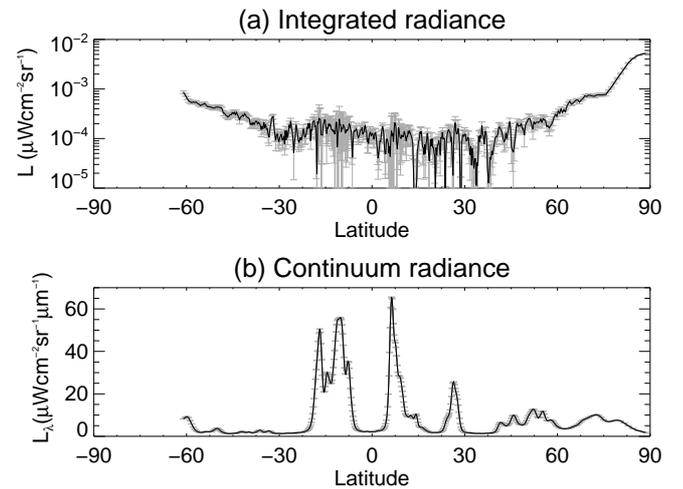}
\caption{(a) The integrated radiance of the H\textsubscript{3}\textsuperscript{+} P(5,5) emission line as a function of latitude. (b) Jupiter's continuum radiance as a function of latitude.} 
\label{fig:latitude}
\end{figure}

\begin{figure*}
\centering
\includegraphics[width=18cm]{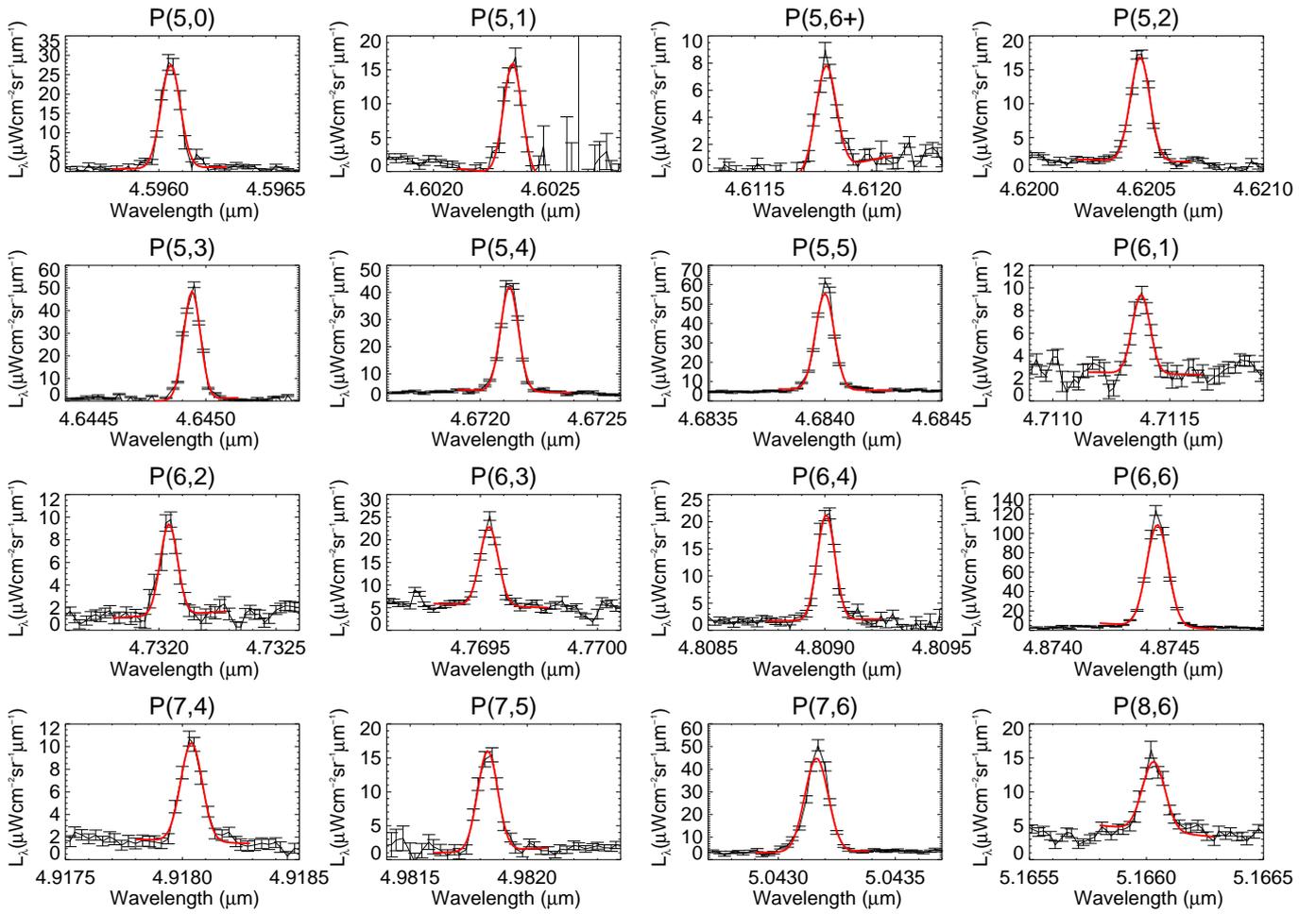}
\caption{Identification of H\textsubscript{3}\textsuperscript{+} emission features in Jupiter's 5-micron window. The data is from the northern polar region (84-86$^{\circ}$N), where the H\textsubscript{3}\textsuperscript{+} signature is particularly strong. Longitudes range from 87$^{\circ}$W to 108$^{\circ}$W.} 
\label{fig:emission_features}
\end{figure*}

Figure ~\ref{fig:latitude}(a) shows how the integrated radiance of the emission line varies with latitude. For comparison, Figure~\ref{fig:latitude}(b) shows the continuum radiance. The strength of the H\textsubscript{3}\textsuperscript{+} emission is a factor of $\sim$100 higher in the polar regions than in the equatorial regions, which is consistent with the fact that H\textsubscript{3}\textsuperscript{+} is predominantly an auroral species. Since the northern polar region has the strongest H\textsubscript{3}\textsuperscript{+} signature, data from this region (84-86$^{\circ}$N) was used to search for additional emission lines. The results of that search are the sixteen lines described in Table~\ref{tab:emission_features}, and they are each shown in Figure~\ref{fig:emission_features}.

\subsection{Kinetic temperature}
\label{sec:kinetic}

The kinetic temperature of the H\textsubscript{3}\textsuperscript{+} ions can be determined by measuring the width of the emission lines. We restricted our analysis to the same northern polar region (84-86$^{\circ}$N) shown in Figure~\ref{fig:emission_features}, where the emission lines are strongest; at other latitudes, the low signal-to-noise prevents reliable results. As in the previous section, we used the MPFIT routine to fit a convolution of a triangle and a Gaussian to the sixteen observed lines. These fits are shown by the red lines in Figure~\ref{fig:emission_features} and the Gaussian FWHM of each line is given in Table~\ref{tab:emission_features}. These Doppler line widths can then then converted into a kinetic temperature, T\textsubscript{kin}, via Equation~\ref{eq:doppler_temp}~\citep[e.g][]{emerson96}:

\begin{equation}
T\textsubscript{kin} = \frac{Mc^2}{8R\ln2} \left(\frac{\Delta\lambda}{\lambda_0}\right)^2
\label{eq:doppler_temp}
\end{equation}

where $M$ is the molar mass of the molecule, $c$ is the speed of light, $R$ is the molar gas constant, $\Delta\lambda$ is the Doppler line width and $\lambda_0$ is the wavelength at the centre of the line. 

If a 10\% error is assumed for the resolving power of the CRIRES instrument, then the average kinetic temperature is 1390$\pm$160 K. This is reasonably consistent with the value of 1150$\pm$60 K that was obtained by~\citet{drossart93} using an emission line at 3.5 \textmu m.

\subsection{Rotational temperature}
\label{sec:rotational}

\begin{figure*}
\centering
\includegraphics[width=14cm]{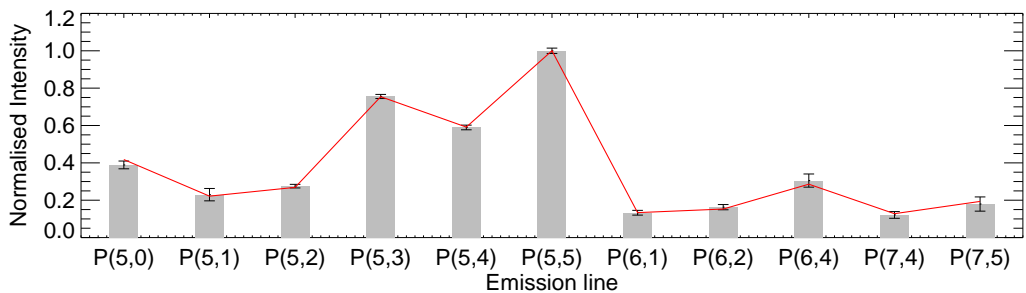}
\caption{Relative intensity of observed H\textsubscript{3}\textsuperscript{+} lines (grey) and the fit obtained with the retrieved T\textsubscript{rot} of 960 K (red). } 
\label{fig:rotational}
\end{figure*}

The rotational temperature of the H\textsubscript{3}\textsuperscript{+} ions can be determined by measuring the relative intensities of different rotational lines within the same vibrational manifold. In this dataset, we have observations of fifteen lines from the same \textnu\textsubscript{2} fundamental band. Since multiple wavelength settings were used to make these observations, we do not have the full set of H\textsubscript{3}\textsuperscript{+} lines at any given spatial location; different lines correspond to different longitudes. However, there is some overlap between the segments, such that several emission lines were observed twice. If we assume that the H\textsubscript{3}\textsuperscript{+} temperature is constant across this region, and that the only variations are in in the H\textsubscript{3}\textsuperscript{+} abundance, then we can scale the different lines so they can be compared. Once scaled, we have three sets of lines that can be directly compared: (i) P(5,0), P(5,1), P(5,2) and P(5,3) at 94$^{\circ}$W (ii) P(5,4), P(5,5), P(6,1) and P(6,2) at 101$^{\circ}$W, (iii) P(6,4), P(7,4) and P(7,5) at 86$^{\circ}$W. Within each set, the H\textsubscript{3}\textsuperscript{+} abundance should be consistent, but the abundance may vary between the sets. The normalised intensities of these lines are shown by the grey bars in Figure~\ref{fig:rotational}.

In the optically thin limit, the integrated radiance of the emission line (relative to the continuum) is proportional to the optical depth, and therefore to the line intensity. As described in~\citet{stallard02}, if LTE is assumed, then the line intensity, $I$, of a particular transition is given by

\begin{equation}
I \propto \frac{\omega g (2J'+1)A}{Q(T\textsubscript{rot})} \exp{\left(-\frac{E'}{kT\textsubscript{rot}}\right)}
\label{eq:rotational}
\end{equation}

where $\omega$ is the wavenumber of the transition, $g$ is the nuclear spin degeneracy factor, $J'$ is the rotational quantum number of the upper state, $A$ is the Einstein A coefficient, $Q(T)$ is the partition function~\citep{miller13}, $E'$ is the energy of the upper state and T\textsubscript{rot} is the rotational temperature.

For each emission line, the parameters $\omega$, $g$, $J'$, $A$ and $E'$ are known~\citep{neale96}. We can therefore search for a temperature $T\textsubscript{rot}$ that reproduces the relative intensities shown in Figure~\ref{fig:rotational}. We used MPFIT to apply a least-squares fit, with four variables: T\textsubscript{rot} and three scaling factors for each of the three sets of emission lines, which are proportional to the H\textsubscript{3}\textsubscript{+} abundance in each case.

This fitting process produced a best-fit T\textsubscript{rot} of 960$\pm$40 K. This fit is shown by the red line in Figure~\ref{fig:rotational}. The retrieved relative abundances are 1.47 (94$^{\circ}$W), 1.00 (101$^{\circ}$W) and 1.26 (86$^{\circ}$W), and the grey bars in Figure~\ref{fig:rotational} have been scaled according to these values. In order to confirm that this fitting process was reliable, the three sets of lines were also fit independently, using just two parameters in each case (T\textsubscript{rot} and a scaling factor). This produced results of 880$\pm$110, 980$\pm$40, 830$\pm$170 K, which is consistent with the overall value of 960$\pm$40 K. This result is also consistent with previously published results for the rotational temperature of H\textsubscript{3}\textsuperscript{+} in Jupiter's atmosphere, which range from 670 K~\citep{oka90} to 1250 K~\citep{drossart93}.

\subsection{Vibrational temperature}
\label{sec:vibrational}

The vibrational temperature of the H\textsubscript{3}\textsuperscript{+} ions can be determined by measuring the relative intensities of emission lines from different vibrational manifolds. In this dataset, we have simultaneous observations of one emission line from an overtone band, P(5,6+), alongside several emission lines from the fundamental band: P(5,0), P(5,1) and P(5,3). Equation~\ref{eq:rotational} can be expressed as

\begin{equation}
T\textsubscript{vib} = \frac{E'_2 - E'_1}{k}\left[\ln{\left(\frac{I_1}{I_2}\frac{(2J'_2+1)\omega_2A_2}{(2J'_1+1)\omega_1A_1}\right)}\right]^{-1}
\end{equation}

where the subscripts 1 and 2 refer to two different emission lines. If we compare the intensity of the overtone line with each of the three fundamental band lines, we obtain vibrational temperatures of 940$\pm$50 K, 910$\pm$80 and 920$\pm$30 K. The mean value of T\textsubscript{vib} is therefore 925$\pm$25 K. Previous studies found T\textsubscript{vib} values of 900-1250 K~\citep{stallard02} and 960$\pm$50~\citep{raynaud04}.

\section{Discussion and conclusions}

In this work, we use high-resolution ground-based observations from the CRIRES instrument at the VLT to identify previously-undetected H\textsubscript{3}\textsuperscript{+} emission lines in Jupiter's 5-micron spectrum: fifteen lines from the \textnu\textsubscript{2} fundamental, and one line from the 2\textnu\textsubscript{2}(2)-\textnu\textsubscript{2} overtone. By considering the broadening and relative intensities of these lines, we measure an average T\textsubscript{kin} of 1390$\pm$160 K, an average T\textsubscript{rot} of 960$\pm$40 K and an average T\textsubscript{vib} of 925$\pm$25 K. All of these values are consistent with previous measurements of H\textsubscript{3}\textsuperscript{+} in Jupiter's ionosphere.

By comparing these temperatures, we can gain insight into whether the assumption of LTE is valid. The Einstein-A coefficients of the rotational transitions are very small, so at the temperatures and densities present in Jupiter's ionosphere, the rotational states are expected to be in LTE~\citep{melin05}. This is turn means that the derived rotational temperature should match the kinetic temperature; we find that T\textsubscript{rot} is slightly lower than T\textsubscript{kin}, which could suggest that this assumption is not valid. The only previous study to compare T\textsubscript{kin} and T\textsubscript{rot} found consistent values \citep[1150K and 1250K,][]{drossart93}. However the T\textsubscript{rot} in this case was larger than the temperature derived by most other studies. It is less surprising that T\textsubscript{vib} is lower than T\textsubscript{kin}, since the theoretical work~\citet{melin05} has shown that vibrational levels depart from LTE in Jupiter's upper atmosphere, and the observational work of~\citet{raynaud04} also found this to be the case.

Departures from LTE will primarily affect the high altitudes (>2000 km above 1 bar), where the atmosphere is both hot and tenuous~\citep{melin05}. This significantly reduces the ability of H\textsubscript{3}\textsuperscript{+} to act as a thermostat in Jupiter's upper atmosphere, as the energy inputs from charged particles and solar radiation are not offset by the cooling provided by H\textsubscript{3}\textsuperscript{+}. In addition, non-LTE effects alter the apparent H\textsubscript{3}\textsuperscript{+} column density, which in turn affects the inferred conductivity of the ionosphere. The degree to which LTE holds is therefore important in understanding the energy budget of the upper atmosphere.

It should be noted that this suggested violation of rotational LTE must await direct confirmation by future measurements.  The key challenge of this study was the use of H\textsubscript{3}\textsuperscript{+} lines in multiple different wavelength settings, observed at different times, meaning that we are potentially convolving different temperatures and ionospheric wind speeds, in addition to different H\textsubscript{3}\textsuperscript{+} abundances. Only an instrument capable of simultaneous observations of multiple H\textsubscript{3}\textsuperscript{+} lines over a broad wavelength range could allow confirmation that T\textsubscript{rot} and T\textsubscript{kin} are genuinely out of equilibrium. This study may be possible with future observations, including IRTF/ISHELL and the refurbished CRIRES.

Previous studies of H\textsubscript{3}\textsuperscript{+} have focussed on the L-band and the K-band. This work opens up a new atmospheric window for future studies: the M-band. This increases the range of instruments that can be used to observe H\textsubscript{3}\textsuperscript{+} auroral emission, and will also allow for simultaneous studies of both the upper and deep atmosphere.

\bibliographystyle{aa}
\bibliography{/Users/rohinigiles/Documents/Work/master.bib}

\end{document}